\documentclass[pra,aps,superscriptaddress, twocolumn]{revtex4}
\usepackage{amsmath,amsfonts,amssymb}
\usepackage{subfigure}
\usepackage{dcolumn}
\usepackage{bm}
\usepackage{tikz}
\usepackage[colorlinks=true,citecolor=blue,urlcolor=blue]{hyperref}
\usepackage{amsmath,amsfonts,amssymb,times,natbib}
\usepackage[standard]{ntheorem}

\begin{document}

\title{Reformulating noncontextuality inequalities in an operational approach}

\author{Zhen-Peng Xu}
 \affiliation{Theoretical Physics Division, Chern Institute of Mathematics, Nankai University,
 Tianjin 300071, People's Republic of China}
 \affiliation{Departamento de F\'isica Aplicada II, Universidad de Sevilla, E-41012 Sevilla, Spain}

\author{Debashis Saha}
 \affiliation{Institute of Theoretical Physics and Astrophysics, National Quantum Information Centre, Faculty of Mathematics, Physics and Informatics, University of Gda\'{n}sk, 80-952 Gda\'{n}sk, Poland}

\author{Hong-Yi Su}
\email{hysu@mail.nankai.edu.cn}
 \affiliation{Theoretical Physics Division, Chern Institute of Mathematics, Nankai University,
 Tianjin 300071, People's Republic of China}
  \affiliation{Department of Physics Education, Chonnam National University, Gwangju 500-757, Republic of Korea}

\author{Marcin Paw\l{}owski}
\email{dokmpa@univ.gda.pl}
 \affiliation{Institute of Theoretical Physics and Astrophysics, National Quantum Information Centre, Faculty of Mathematics, Physics and Informatics, University of Gda\'{n}sk, 80-952 Gda\'{n}sk, Poland}

\author{Jing-Ling Chen}
\email{chenjl@nankai.edu.cn}
 \affiliation{Theoretical Physics Division, Chern Institute of Mathematics, Nankai University,
 Tianjin 300071, People's Republic of China}
 \affiliation{Centre for Quantum Technologies, National University of Singapore,
 3 Science Drive 2, Singapore 117543}


\begin{abstract}
A new theory-independent noncontextuality inequality is presented [Phys. Rev. Lett. 115, 110403 (2015)] based on Kochen-Specker (KS) set without imposing the assumption of determinism. By proposing novel noncontextuality inequalities, we show that such result can be generalized from KS set to the noncontextuality inequalities not only for state-independent but also for state-dependent scenario. The YO-13 ray and $n$ cycle ray are considered as examples.
\end{abstract}


\maketitle

\section{Introduction}
``Quantum contextuality" is one of the most intriguing and fundamental features of quantum mechanics. The fact that a \textit{deterministic noncontextual} ontic theory cannot reproduce all the operational predictions of  quantum theory was first pointed out by Bell-Kochen-Specker \cite{Specker60,Bell66,KS67,Mermin93}. In the noncontextual model considered by Kochen-Specker (KS), a predefined value 0 or 1 is assigned to every projector  for any ontic state in a context independent way such that the assigned value is same irrespective of the other projectors appear in the observable. The KS argument to show contextuality is based on the incompatibility of such noncontextual model for a set of projectors. A set of projectors with this property is known as KS set. On the other hand, state-independent contextuality (SIC) is based on an inequality involving experimentally testable quantity  which is satisfied by all noncontextual models and violated by any quantum state of a particular dimension. SIC is a general approach to test contextuality than KS set and has been considerably studied in recent years \cite{Cabello08,badziag2009,yu2012}.

While the assumption of determinism in the argument of KS contextuality is arguable \cite{spekkens2014}, it can be inferred from another notion of noncontextuality, namely, \textit{preparation noncontextuality for mixed states} introduced by Spekkens \cite{spekkens2005}. If different preparation procedures for a mixed state cannot be distinguished operationally, then, in a preparation noncontextual model, the ontic description for those preparations is the same. The assumption of preparation noncontextuality together with the predictions of quantum theory that measurement outcomes of orthogonal pure states are perfectly predictable (when they form the eigen basis of that measurement) implies outcome determinism in sharp measurements. Therefore, from a KS contextuality test, one may exclude noncontextual models if the outcome determinism is justified by the assumption of preparation noncontextuality. But to do so, one has to separately verify the \textit{perfect predictability} condition, which itself cannot be precisely implemented in a real experiment.

To address this fundamental issue, an alternative noncontextuality inequality is proposed by Kunjwal-Spekkens \cite{kunjwal2015}. The inequality is derived precisely assuming preparation and measurement noncontextual model in an operational approach and is based on Cabello-Estebaranz-GarciaAlcaine (CEG) 18-ray \cite{cabello1999,Cabello08} which is the minimal KS set in dimension four. An experiment \cite{mazurek2015} has been performed according to this method. We refer to such inequalities as operational noncontextuality inequalities (ONCI).

In this work, we extend the ONCI from KS set to state-independent as well as state-dependent KS noncontextuality inequalities. First, a general method is described to obtain ONCI based on any noncontextuality inequalities using the constraints from the operational predictions of those inequalities.  Since YO-13 ray \cite{yu2012}  and KCBS ray \cite{klyachko2008} are the minimal sets for SIC and state-dependent contextuality in three-dimensional system, we take these scenarios for explicit illustration. Later, we propose a efficient ONCI in terms of the robustness of quantum violation without the prior form KS noncontextuality inequalities. The inequality is derived for YO-13 ray and arbitrary $n$ cycle ray \cite{LSW,AQBCC} and it's quantum violation is studied.\\

\section{Noncontextual ontological model}
Let $\mathcal{P},\mathcal{M}$ be the sets of all preparations and measurements respectively in an operational theory, and $p(k|M,P)$ be the observed probability of getting an outcome $k\in\mathcal{K}_M$ after a measurement procedure $M\in \mathcal{M}$ on a preparation $P \in \mathcal{P}$.

We call $\{\Lambda, \mu, \xi\}$ an ontological model of the operational theory $\{\mathcal{P}, \mathcal{M}, p\}$. $\Lambda$ is the ontic state space (also known as hidden-variable space) and $\mu(\lambda|P) \in [0,1]$ is the probability distribution over the ontic space for some preparation $P$, and $\xi(k|M,\lambda)$ denotes response function for the outcome $k$ of the measurement $M$ when the ontic state of the system is $\lambda$. Since the ontological model should reproduce the predictions of operational theory, we have $\forall k\in\mathcal{K}_M, M\in \mathcal{M}, P\in\mathcal{P}$,
\begin{equation}\begin{split}
& \sum_{\lambda\in\Lambda} \mu(\lambda|P) = 1, \\
& p(k|M,P) = \sum_{\lambda\in\Lambda} \mu(\lambda|P) \xi(k|M,\lambda).
\end{split}\end{equation}
We denote the event of obtaining an outcome $k$ for measurement $M$ by $[k|M]$. Two events $[k|M]$ and $[k'|M']$ are said to be operationally equivalent if
\[
p(k|M,P) = p(k'|M',P), \ \forall P \in \mathcal{P}.
\]
This is denoted by as $[k|M]\approx [k'|M']$. An ontic-theory is said to be {\it measurement-noncontextual} if
\begin{equation}\begin{split}\label{mnc}
[k|M]\approx [k'|M'] \Rightarrow \xi(k|M,\lambda) = \xi(k'|M',\lambda),\ \forall \lambda\in\Lambda.
\end{split}\end{equation}

Two different preparation procedures $P,P'$ of mixed states (i.e. corresponding to a same density matrix) are operationally equivalent ($P\approx P'$) in the sense that
\[
p(k|M,P) = p(k|M,P'),\ \forall k\in\mathcal{K}_M, M\in \mathcal{M}.
\]
Similarly an ontological model is said to be {\it preparation-noncontextual for mixed states} if
\begin{equation}\begin{split}\label{pnc}
 P\approx P' \Rightarrow \mu(\lambda|P) = \mu(\lambda|P'),\ \forall \lambda\in\Lambda.
\end{split}\end{equation}

KS theorem states that \textit{when $M$ is sharp measurement and $\xi(k|M,\lambda) \in [0,1]$ is deterministic, a measurement noncontextual ontic model (\ref{mnc}) cannot reproduce all quantum mechanical predictions}. More generally, one can verify quantum contextuality from the violation of an inequality, in terms of experimentally measurable quantity, which is satisfied by KS noncontextual model. A standard form of such inequality
\begin{equation}\label{I}
\mathcal{I} = \sum_{i} c_i p(P_i|P) \leqslant \alpha
\end{equation}
is true for all KS noncontextual theories, where $P$ is a preparation, $P_i$'s are projectors appearing in some measurement, $p(P_i|P)$ denotes the probability of getting the measurement outcome which corresponds to the projector $P_i$. If the quantum predictions of $\mathcal{I}$ is greater than  $\alpha$, for all possible preparations $P$, then the noncontextuality inequality is known as SIC inequality. On the other hand, state-dependent contextuality implies the violation of $\mathcal{I}$ for some preparation. For every KS contextuality argument, one can always associate a graph  where each vertex is a projector and two projectors are orthogonal to each other if the vertices are adjacent. \\

\section{Review of operational noncontextual inequality}
The objective is to derive a theory-independent noncontextuality inequality using only the assumptions of \eqref{mnc} and \eqref{pnc}. One can assign a measurement $M_i$ corresponds to each $d$-clique in the KS graph, where $d$ is the dimension of the system. A preparation $P_{i,k}$ is such that the event $[k|M_i]$ is certain and  the effective preparation $P_i^{\text{ave}}$ is obtained by sampling all possible $k$ uniformly at random and then implementing $P_{i,k}$. Then the inequality
\begin{equation}\label{generala}
\mathcal{A} = \frac{1}{cd}\sum_{i=1}^c \sum_{k=1}^d p(k|M_i,P_{i,k}) \leqslant a
\end{equation}
is valid in all ontic theories satisfying \eqref{mnc} and \eqref{pnc}, where $c$ is the number of $d$-cliques in the KS graph. The value of $a$ is less than 1, whereas it can be readily checked that the operational prediction of $\mathcal{A}$ in quantum theory is 1. The inequality proposed in \cite{kunjwal2015} is based on a KS set known as BEC-18 ray in four dimensional system which possesses a particular property that all the vertices appear in $d$-cliques twice. In general, KS set does not satisfy this condition and subsequently the inequality $(\ref{generala})$ cannot be trivially generalized for any KS noncontextuality inequality \eqref{I}.\\

\section{Operational noncontextuality inequality based on the original inequality}

Before we outline the method to derive a ONCI corresponds to any KS noncontextuality inequality, let's first demonstrate it for the YO-13 ray \cite{yu2012} in SIC and KCBS ray \cite{klyachko2008} in state-dependent scenario.

\subsection{YO-13 ray}
\begin{figure}
\centering
\includegraphics{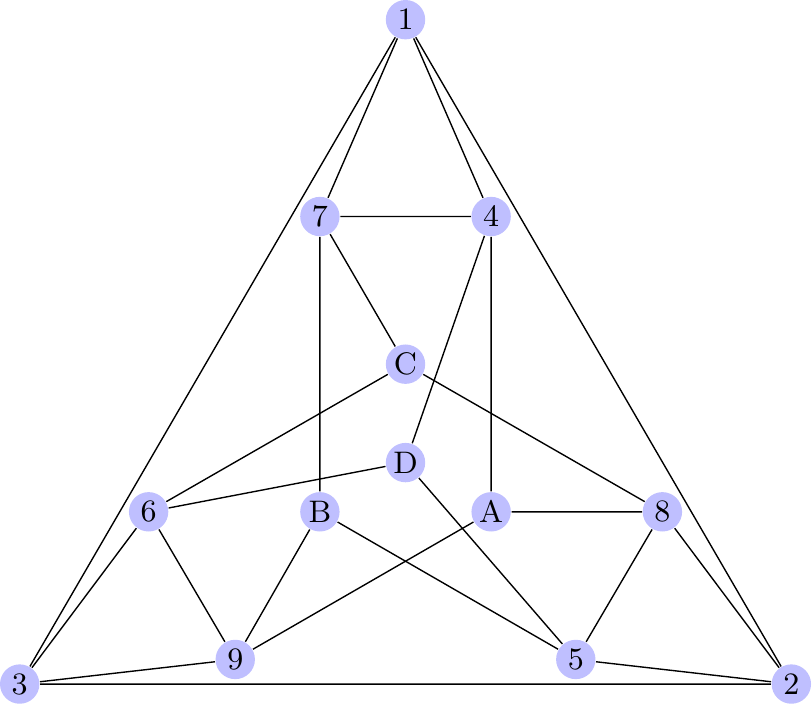}
\caption{The YO-13-ray graph.}\label{yo13ray}
\end{figure}

The SIC inequality based on the YO-13 ray
\begin{equation}\label{yoineq}
\mathcal{I} = 2\sum_{i=1}^9 p(P_i|P) + \sum_{i=A,B,C,D} p(P_i|P) \leqslant 7
\end{equation}
holds for KS noncontextual theories, while $\mathcal{I} = \frac{22}{3}$ in quantum theory.
Let's consider the YO-13 graph shown in Fig. \ref{yo13ray}. A clique is said to be maximal if it's not contained in any larger one. There are $16$ maximal cliques $C_i,i\in \{1,\ldots,16\}$ in YO-13-ray:
\begin{eqnarray}\label{mc}
&(1,2,3),(1,4,7),(2,5,8),(3,6,9),\nonumber \\
&(4,A),(4,D),(5,B),(5,D),(6,C),(6,D),\\
&(7,B),(7,C),(8,A),(8,C),(9,A),(9,B).\nonumber
\end{eqnarray}
Let $M_i\in\mathcal{M}:i\in\{1,\ldots,16\}$ be $16$ three-outcome measurements correspond to all the cliques and $P_{i,k} \in \mathcal{P},i\in\{1,2,...,16\},k\in\{1,2,3\}$ be $48$ preparation procedures. We associate the measurement $M_i$ with clique $C_i$, and the preparation procedure $P_{i,k}$ with the $k$-th projector in measurement $M_i$.

Consider the effective preparation $P_i^{\text{ave}}$ defined as the procedure obtained by sampling $k$ uniformly at random and implementing $P_{i,k}$. It can be checked that $P_1^{\text{ave}} \approx P_{2}^{\text{ave}} \approx P_{3}^{\text{ave}} \approx P_{4}^{\text{ave}}$, which is eventually the preparation of maximally mixed state, denoted by $P_{m}$. So preparation-noncontextuality for mixed states \eqref{pnc} implies, for $i\in\{1,2,3,4\}$, that
\begin{equation}\label{pnc13}
\mu(\lambda|P_m) = \frac{1}{3} \sum_k \mu(\lambda|P_{i,k}), \forall \lambda \in \Lambda.
\end{equation}
It can also be verified that two events $[k|M_i]\approx [k|M_{i'}]$ when they are associated with the same vertices. By assuming measurement noncontextuality \eqref{mnc}, given an ontic state $\lambda$, we can assign nonnegative values, say $w_1,...,w_9,w_A,w_B,w_C,w_D$, for each of the vertices, for example,
\begin{equation}\begin{split}
&\xi(1|M_1,\lambda) = \xi(1|M_2,\lambda) = w_1,\\
&\xi(2|M_5,\lambda) = \xi(2|M_{13},\lambda) =  \xi(2|M_{15},\lambda)  = w_A,
\end{split} \end{equation} and so on. Thus for a given ontic state, we have a space of response functions with 13 variables. Each of the 16 cliques (\ref{mc}) yields a constraints on that space:
\begin{equation}\begin{split}
&w_1+w_2+w_{3}=w_1+w_4+w_{7}=w_2+w_{5}+w_{6}\\
&=w_{3}+w_{6}+w_{9}=1,
\end{split}\end{equation}
and
\begin{equation}\begin{split}
&(w_4+w_A),(w_4+w_D),(w_5+w_B),(w_5+w_D),(w_6+w_C),\\
&(w_6+w_D), (w_7+w_B),(w_7+w_C),(w_8+w_{A}),(w_8+w_C),\\
&(w_9+w_A),(w_9+w_{B}) \leq 1.
\end{split}\end{equation}   Under these constraints the extremal points of the space containing $w$-s (response function space) can be obtained. In this case, there are 420 such points.

As mentioned earlier, we consider the following quantity
\begin{equation}\label{yo13a}
\mathcal{A} = \frac{1}{9}\sum_{i=2}^4\sum_{k=1}^3 p(k|M_i,P_{i,k}).
\end{equation}
One can check that sum of projectors corresponds to the four preparations $P_{5,2},P_{7,2},P_{9,2},P_{10,2}$ is $\frac{4}{3} \mathbb{I}$ and therefore $\forall P\in\mathcal{P}$, $\sum_{i=5,7,9,10} p(2|M_i,P) =\frac{4}{3}$. Let's also define $\Lambda(a)$ is a subset  of $\Lambda$ such that
\[
\sum_{k\in \{5,7,9,10\}} \xi(2|M_k,\lambda) \geqslant a,\ \forall \lambda\in\Lambda(a),
\]
and $m_P(S) = \sum_{\lambda\in S} \mu(\lambda|P), \forall S\subset \Lambda$. Then,
\begin{equation}\label{yoabcd}
\begin{split}
& \sum_{i=5,7,9,10} p(2|M_i,P) = \sum_\lambda \sum_{k\in \{5,7,9,10\}} \xi(2|M_k,\lambda) \mu(\lambda|P)\\
& = \sum_{\lambda\in \Lambda(a)} \sum_{k\in \{5,7,9,10\}} \xi(2|M_k,\lambda) \mu(\lambda|P) \\
& + \sum_{\lambda\in \Lambda-\Lambda(a)} \sum_{k\in \{5,7,9,10\}} \xi(2|M_k,\lambda) \mu(\lambda|P)\\
&\leqslant  \frac{8}{3} m_P(\Lambda(a)) + a (1-m_P(\Lambda(a))),
\end{split}
\end{equation}
where $a\in [1,\frac{4}{3}]$,  Note that the above relations holds because $w_A + w_B + w_C + w_D \leqslant 8/3$ which is verified by checking all the extremal points of the response function space.

By denoting $m_P(S)$ as $m(S)$ for $P=P_m$, which is the maximally mixed state, Eq. \eqref{yoabcd} implies $m(\Lambda(a)) \geqslant (4/3-a)/(8/3-a)$. Further the same analysis shows that
\begin{eqnarray}\label{AA(a)}
\mathcal{A} \leqslant (1-m(\Lambda(a))) + \mathcal{A}(a) m(\Lambda(a)),
\end{eqnarray}
where
$$
\mathcal{A}(a) = \frac{1}{m(\Lambda(a))}\sum_{\lambda\in \Lambda(a)} \frac{1}{9}\sum_{i=2}^4\sum_{k=1}^3 \xi(k|M_i,\lambda) \mu(\lambda|P_{i,k}).
$$
Now we want to have an upper bound on $\mathcal{A}(a)$. Clearly, the upper bound of $\mathcal{A}(a)$ is a convex function in the area $\Lambda(a)$. To prove that the upper bound of $\mathcal{A}(a)$ is a linear function of $a\in [1,4/3]$, it suffices to show that the upper bounds of $\mathcal{A}(1),\mathcal{A}(a'),\mathcal{A}(4/3)$ lie on the same line, taking $a' \in (1,4/3)$. To calculate the upper bound of $\mathcal{A}(a)$, one can generate all the extremal points of the response function space by imposing the additional constraints, $1 \leqslant w_A + w_B + w_C + w_D \leqslant 4/3$. It is first observed that the upper bound of $\mathcal{A}(7/6) = 17/18$, and further conclude $\mathcal{A}(a) \leqslant (4-a)/3, \forall a\in [1,4/3]$. Substituting this relation to \eqref{AA(a)}, we obtain $\mathcal{A} \leqslant 4\sqrt{5}/9 \approx 0.994$. By construction the quantum prediction of $\mathcal{A}$ is 1. \\

\subsection{KCBS ray}
\begin{figure}[http!]
  \centering
  \subfigure[\hspace{-1em}]{
    \label{kcbsa} 
    \includegraphics{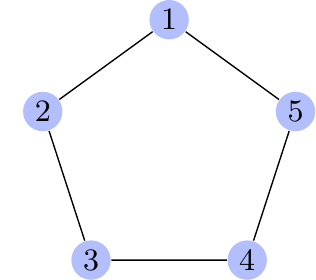}}
  \subfigure[\hspace{-1em}]{
    \label{kcbsb} 
    \includegraphics{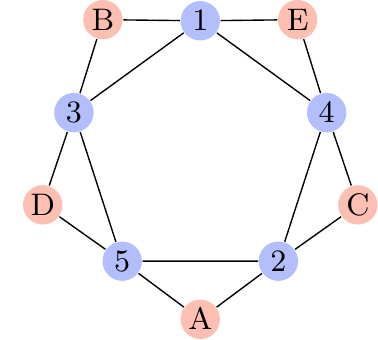}}
  \caption{The original KCBS graph (a) and the one (b) after adding new vertices $A,B,C,D,E$.}
  \label{newkcbs} 
\end{figure}

The KCBS inequality is given by
\[
\mathcal{I} = \sum_{i=1}^5 p(P_i|P) \leqslant 2,
\]
where two measurement $P_i, P_j$ are exclusive if vertices $i,j$ are adjacent in the graph shown in Fig. \ref{kcbsa}. While $\sqrt{5}$ is the maximal value that quantum theory can achieve with a $3$ dimensional system.
Since the quantum violation is state-dependent, the first step is to add new vertices to the original graph such that each edge is included in a clique containing $3$ vertices. The final graph is Fig. \ref{kcbsb} and the $5$ cliques $C_i$ are
\begin{equation}\label{mc5}
(1,2,A),(2,3,B),(3,4,C),(4,5,D),(5,1,E).
\end{equation}

Let $M_i\in\mathcal{M}:i\in\{1,\ldots,5\}$ corresponding to the clique $C_i$ be $5$ three-outcome projective measurements and $P_{i,k} \in \mathcal{P},(i\in\{1,\ldots,5\},k\in\{1,2,3\})$ be $15$ preparation procedures. Similarly, the preparation procedure $P_{i,k}$ is associated with the $k$-th vertices in clique $C_i$. It can be checked that two events $[k|M_i]\approx [k|M_{i'}]$ when they are associated with the same vertices and $P_1^{\text{ave}} \approx \cdots \approx P_{5}^{\text{ave}}\approx P_m$, where the effective preparation $P_i^{\text{ave}}$ is defined as the procedure obtained by sampling $k$ uniformly at random and implementing $P_{i,k}$.
Further assuming measurement noncontextuality \eqref{mnc}, we construct a space of response function of 10 variables, denoted by $w_i, i \in \{1,...,5,A,B,C,D,E\}$. By imposing five equality constraints from five cliques, we can obtain all the 12 extremal points of that space.

Now we want to derive a noncontextual bound of the following quantity
\begin{equation}
\mathcal{A} = \frac{1}{15}\sum_{i=1}^5 \sum_{k=1}^3 p(k|M_i,P_{i,k}).
\end{equation}
under the constraint that $\exists P \in \mathcal{P}$ is  such that
\begin{equation}\label{kcbscon}
\sum_{i=1}^5 p(1|M_i,P) = \sqrt{5}.
\end{equation}
Similar analysis using the extremal points, as in the YO-13 ray case, the Eq. \eqref{kcbscon} implies that
\begin{eqnarray}
\sqrt{5} \leqslant \frac{5}{2} m_P(\Lambda(a)) + a (1-m_P(\Lambda(a))),
\end{eqnarray}
where $a \in [2,\sqrt{5}]$ and $\Lambda(a)$ is such a subset of $\Lambda$ that
\[
\sum_{i=1}^5 \xi(1|M_i,\lambda) \geqslant a,\ \forall \lambda\in\Lambda(a).
\]
On the other hand,
\begin{eqnarray*}
\mathcal{A} &\leq & (1-m(\Lambda(a))) + \mathcal{A}(a) m(\Lambda(a)),
\end{eqnarray*}
where
\[
\mathcal{A}(a) = \frac{1}{m(\Lambda(a))} \sum_{\lambda\in\Lambda(a)}\frac{1}{5} \sum_{i=1}^5 \xi(M_i,\lambda) \mu(\lambda|P_{i,k}).
\]
To get an upper bound on $\mathcal{A}(a)$ we follow the same method described earlier. We generate all the extremal points imposing the additional constraint, $2 \leqslant w_1+w_2+w_3+w_4+w_5 \leqslant \sqrt{5}$. Then taking an intermediate point, it is obtained that $\mathcal{A}(a) \leqslant \frac{17-6a}{5}$. And it's easy to see that $m(\Lambda(a)) \geqslant m_{P}(\Lambda(a))/3$. Thus, if we choose $a=\frac{1}{2} \left(5-\sqrt{5-2 \sqrt{5}}\right) \approx 2.13$, then
\[
\mathcal{A} \leqslant 1-\frac{2}{5} \left(\sqrt{5}+\sqrt{5-2 \sqrt{5}}-3\right) \approx 0.985.
\] \\

The general scheme to derive a nontrivial bound of the quantity $\mathcal{A}$ given in \eqref{generala} is as follows. In the state-dependent scenario, new vertices are added in the original graph corresponding to the KS noncontextuality inequality such that each edge is in a clique of $d$ vertices, where $d$ is the dimension of the system.
Then we estimate the measure of the set of ontic states $\lambda$'s that violate the noncontextual bound of the original inequality $\mathcal{I}$. Further the operational prediction of the original KS noncontextuality inequality is used as an additional constraint to obtain an upper bound of $\mathcal{A}$ on the set of ontic states. This implies a nontrivial bound over the observed quantity $\mathcal{A}$. However the magnitude of quantum violation is very low. To address this problem we propose a modified inequality in the subsequent section.\\


\subsection{Efficient operational noncontextuality inequality}
\begin{figure}[http!]
  \centering
  \includegraphics{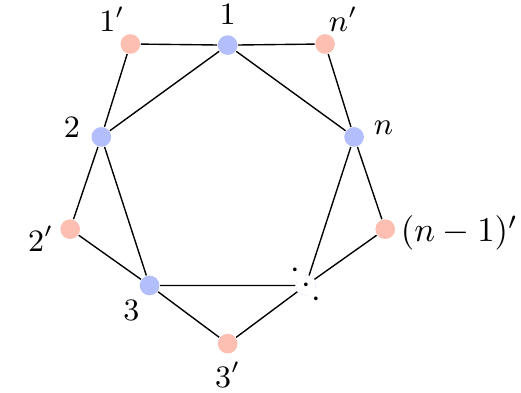}
  \caption{The $n$ cycle graph after adding new vertices $1',2',\ldots,n'$.}
  \label{newchain} 
\end{figure}

To optimize the quantum violation, here we propose an elegant inequality which is a linear combination of $\mathcal{A}$ and an additional part involving the projectors which do not appear in the $d$-cliques of the KS graph.
For the YO-13 ray, using \eqref{pnc13} and by involving all the extremal points of response function space, one can obtain the following relation
\begin{widetext}
\begin{equation}
\begin{split}
\mathcal{A'} &= 3 \mathcal{A} +  \sum_{k\in \{5,7,9,10\}} p(2|M_k,P_m)
 =  \frac{1}{3}\sum^{4}_{i=2}\sum^{3}_{k=1}\sum_{\lambda} \xi(k|M_i,\lambda) \mu(\lambda|P_{ik})
 + \sum_{k\in \{5,7,9,10\}} p(2|M_k,P_m) \\
& \leqslant  \sum_{\lambda}  \sum_{i=2}^4  \eta(M_i,\lambda) \sum^{3}_{k=1} \frac{1}{3} \mu(\lambda|P_{ik})  +\sum_{\substack{k\in \{5, \\ 7,9,10\}}}  \xi(2|M_k,\lambda) \mu(\lambda|P_m)
\;\;\;\;\;\;\left[\text{where $\eta(M_i,\lambda) = \max_{k=1,2,3} \xi(k|M_i,\lambda)$}\right] \\
& =  \sum_{\lambda} \left( \sum_{i=2}^4  \eta(M_i,\lambda)  + \sum_{k\in \{5,7,9,10\}}  \xi(2|M_k,\lambda) \right) \mu(\lambda|P_m)\\
& \leqslant  \max\{w_1,w_4,w_7\} +  \max\{w_2,w_5,w_8\}
 + \max\{w_3,w_6,w_9\}
+ w_A+w_B+w_C+w_D \leqslant 4.
\end{split}
\end{equation}
While in the quantum case the value of the above quantity is $\frac{13}{3}$. Thus, the quantum to classical ratio is $\frac{13}{12}$ which is greater than that of the original KS noncontextuality inequality \eqref{yoineq}, and remarkably equal to maximal violation found in \cite{KBLGC} with context size two.

In the state-dependent scenario, we consider a general KS noncontextuality inequality with a set of $n$ projectors which comprises $n$ cycle KS graph \cite{LSW,AQBCC}. Let's denote each of the vertices of the $n$ cycle KS graph by $i$ and add new vertices $i'$ ($i,i' \in \{1,...,n\}$) such that $(i,{i+1},i')$ is three-clique that corresponds to the measurement $M_i$ (see Fig. \ref{newchain}). Note that the sum of the indexes is taken to be modulo $n$. We also define the preparation $\bar{P}$ by equal mixture of two quantum states orthogonal to a preparation $P$ of a three dimensional system, such that $\frac{1}{3}P + \frac{2}{3}\bar{P} = P_m$. Since $P$ and $\bar{P}$ are perfectly distinguishable, the support of these two preparation in the $\Lambda$ space is disjoint. Similarly, one can have the response function space of $2n$ variables $w_i,w_i'$. The following quantity
\begin{equation}
\begin{split}
\mathcal{A'} &= 3\mathcal{A} + \mathcal{I}
= \frac{1}{n}\sum^{n}_{i=1}\sum^{3}_{k=1} p(k|M_i,P_{i,k}) + \sum^{n}_{i=1} p(1|M_i,P)\\
&\leqslant \frac{3}{n}\sum^{n}_{i=1} \sum_{\lambda} \eta(M_i,\lambda) \mu(\lambda|P_m) + \sum_{\lambda} \sum^{n}_{i=1} \xi(1|M_i,\lambda) \mu(\lambda|P)
\;\;\;\;\;\;\;\;\; \left[\text{ where $\eta(M_i,\lambda) = \max_{k=1,2,3} \xi(k|M_i,\lambda)$} \right]\\
&= \frac{1}{n} \sum_{\lambda \in \Lambda(P)} \left( \sum^{n}_{i=1} \eta(M_i,\lambda) + n \sum^{n}_{i=1} \xi(1|M_i,\lambda)\right) \mu(\lambda|P)
+ \frac{2}{n} \sum_{\lambda \in \Lambda(\bar{P})} \sum^{n}_{i=1} \eta(M_i,\lambda) \mu(\lambda|\bar{P})\\
&\leqslant  \frac{1}{n} \sum^{n}_{i=1} \max\{w_i,w_{i+1},w_{i'}\} + \sum^{n}_{j=1} w_j + 2 .
\end{split}\end{equation}

\end{widetext}
It is obvious that the maximum bound of $\mathcal{A'}$ is attained when $w_{i'}=0, \forall i'$. Thus we need to consider the extremal points of the response function space $w_i$s such that $w_i+w_{i+1} \leqslant 1$. It is shown in \cite{Padberg} that the extremal points of $w_i$ under such constraints can have values from $\{0,\frac{1}{2},1\}$. Using this fact, a simple calculation gives the upper bound, $\mathcal{A'} \leqslant \lfloor \frac{n}{2} \rfloor + 3$. From the result given in \cite{LSW,AQBCC}, we know there exists preparation $P$ and projectors which provide the maximal quantum violation $3+ \frac{n\cos{\frac{\pi}{n}}}{1+\cos \frac{\pi}{n}}$ for odd $n$, and $3+\frac{n}{2}$ for even $n$.\\

\section{Conclusion}
Given a violation of KS noncontextuality inequality, whether one can truly discard noncontextual model or defend noncontextuality by abandoning determinism, is a fundamental concern.
In this work, we provide a solution to this question by extending KS noncontextuality inequalities to an operational noncontextuality inequalities without imposing determinism.

An open question is to find the optimal ONCI given a particular KS set of projectors. Since the bounds of different theories in KS contextuality are related to graph theoretic quantity \cite{CSW}, it would be interesting to look at ONCI in a graph theoretic approach.

\acknowledgments
We thank Jin-Shi Xu for discussions. J.L.C. is supported by the National Basic Research Program (973 Program) of China under Grant No.\ 2012CB921900 and the Natural Science Foundations of China (Grant Nos.\ 11175089 and 11475089). D.S. acknowledges the NCN grant 2014/14/E/ST2/00020 for support. This work is also supported by Institute for Information and Communications Technology Promotion (IITP) grant funded by the Korea Government (MSIP) (No. R0190-16-2028, Practical and Secure Quantum Key Distribution).

Z.P.X. and D.S. contributed equally to this work.

\end{document}